\documentclass[%
 aip,
 amsmath,amssymb,
 reprint,%
]{revtex4-1}

\usepackage{graphicx}
\usepackage{dcolumn}
\usepackage{bm}

\usepackage[utf8]{inputenc}
\usepackage[T1]{fontenc}
\usepackage{mathptmx}
\usepackage{etoolbox}

\makeatletter
\def\@email#1#2{%
 \endgroup
 \patchcmd{\titleblock@produce}
  {\frontmatter@RRAPformat}
  {\frontmatter@RRAPformat{\produce@RRAP{*#1\href{mailto:#2}{#2}}}\frontmatter@RRAPformat}
  {}{}
}%
\makeatother
\begin{document}

\newcommand*{\citen}[1]{%
  \begingroup
    \romannumeral-`\x 
    \setcitestyle{numbers}%
    \cite{#1}%
  \endgroup   
}

\preprint{AIP/123-QED}

\title{Comparison of integral equation theories of the liquid state}

\author{Ilian Pihlajamaa}
\affiliation{%
Soft Matter \& Biological Physics, Department of Applied Physics,
Eindhoven University of Technology, P.O. Box 513, 5600MB Eindhoven, The Netherlands
}%
\author{Liesbeth M.C. Janssen}%
 \email{l.m.c.janssen@tue.nl}
\affiliation{%
Soft Matter \& Biological Physics, Department of Applied Physics,
Eindhoven University of Technology, P.O. Box 513, 5600MB Eindhoven, The Netherlands
}%

\date{\today}

\begin{abstract}
The Ornstein-Zernike equation is a powerful tool in liquid state theory for predicting structural and thermodynamic properties of fluids. Combined with a suitable closure, it has been shown to reproduce \textit{e.g}.~the static structure factor, pressure, and compressibility of liquids to a great degree of accuracy. However, out of the multitude of closures that exist for the Ornstein-Zernike equation, it is hard to predict \textit{a priori} which closure will give the most accurate predictions for the system at hand. To alleviate this problem, we compare the predictive power of many closures on a curated set of representative benchmark systems, including those with hard-sphere, inverse power-law, Gaussian core, and Lennard-Jones particles, in three and two dimensions. For example, we find that the well-known and highly used Percus-Yevick closure gives significantly worse predictions than lesser-known closures of equal complexity in all cases studied. 
We anticipate that the trends observed in our results will aid in making more informed decisions regarding closure choices. To facilitate the adoption of more modern closure theories, we also have packaged, documented, and distributed the code necessary to numerically solve the equations for a given closure and pair interaction potential.
\end{abstract}

\maketitle


\section{Introduction}

Integral equation theories are powerful tools for elucidating the structural and thermodynamic properties of simple liquids \cite{hansen2013theory, solana2013perturbation}. In particular, the conception of the Ornstein-Zernike equation and associated closure theories has facilitated the establishment of a powerful link between particle interactions and equilibrium structural correlation functions. Given such correlations, one can predict a myriad of thermodynamic properties, including potential energy, pressure, compressibility, and free energy \cite{hansen2013theory}.

In addition, the integral equation theories based on the Ornstein-Zernike equation have been successfully adapted to a plethora of other applications, such as nonspherical particles \cite{perera1987solution, blum1972invariant}, molecular systems \cite{chen1971statistical, chandler1972optimized} and polymeric fluids \cite{benmouna1988application, hoye2003self, hoye2004ornstein}, inhomogeneous fluids \cite{henderson1988pair, mccoy1989theory}, lattice models \cite{parola1985critical, stell1969some}, quantum systems \cite{shinoda2001generalized, attard2019quantum, shinoda2001molecular}, percolation problems \cite{coniglio1977pair, chatterjee2000continuum}, and non-equilibrium liquids \cite{brader2013nonequilibrium, eu2006dynamic, franz2015quasi}. Moreover, the theory can be applied inversely \cite{march1984thermodynamically, heinen2018calculating}, i.e.~to answer the question: ``What interaction potential produces the given correlation function at the given state point?''

All of the above applications have in common that they require a closure relation to solve the equations in question. Over the years, many such theories have been constructed (see \citet*{bomont2008recent} for a review) and they have been widely applied to great success. Still, the closure relations remain approximate, leading to occasional, and sometimes substantial, discrepancies compared to simulation results, especially at high densities or low temperatures \cite{lee1995accurate}. Even in the intermediate density regime, the choice of closure can significantly impact predictive accuracy. The abundance of available closure options further complicates this selection process, rendering it a formidable task. In practice, a handful of well-established closures tend to dominate usage not because they exhibit superior accuracy or thermodynamic consistency, but often due to their familiarity or ease of numerical implementation. Indeed, for many common model systems, it is unknown how most closures perform. This is exacerbated by the fact that newly developed closure relations are typically tested on or constructed for one or a few specific reference systems in three dimensions, leaving room for uncertainty about their generalizability.

In this work, we aim to provide guidance for choosing the right closure for a specific system of interest. Our approach involves a systematic exploration of the theory's performance across a diverse spectrum of closures for several systems. While an exhaustive treatise is unfeasible, we have curated a set of benchmark liquids including the hard-sphere, inverse power-law, Gaussian-core, and Lennard-Jones liquids, each in two and three dimensions at multiple state points. These systems were chosen to cover hard impenetrable, soft impenetrable, penetrable, and attractive particles. Comparing the predicted correlation functions with particle-based Monte Carlo simulations, we test every model fluid with a list of closure relations, spanning from the well-known Percus-Yevick closure \cite{percus1958analysis} to more contemporary ones. We focus our comparison on correlation functions over thermodynamic properties because accurate predictions of the correlation functions imply that the predicted thermodynamic properties are accurate as well, while the converse is not necessarily true. To maintain brevity, in this study, our emphasis is placed on uncharged homogeneous liquids that do not comprise multiple components. 

To facilitate the broader utility and reproducibility of our research, we have documented, packaged, and openly published the software tools required for solving the integral equations~\cite{OZjl}. Thus, it should be straightforward not only to reproduce our results but also to apply our methods to new systems of interest. 

\section{Theory}

\subsection{Liquid state theory}

In a homogeneous system of particles that interact with a spherically symmetric pair interaction potential, the Ornstein-Zernike relation \cite{ornstein1914accidental, hansen2013theory}
\begin{equation}\label{eq:oz}
    h(r) = c(r) + \rho \int \mathrm{d} \textbf{r}' c(\textbf{r}-\textbf{r}') h(\textbf{r}')
\end{equation}
relates the total correlation function $h(r) = g(r) - 1$, to the direct correlation function $c(r)$. The radial distribution function, denoted as $g(r)$, quantifies the probability of encountering two particles at a distance described by the vector $\textbf{r}=\textbf{r}_2-\textbf{r}_1$. Because the system is homogeneous and the pair interactions are central, such correlations depend only on the length $r=|\textbf{r}|$. In Eq.~\eqref{eq:oz}, $\rho = N/V$ is the number density of the system, with $N$ and $V$ being the number of particles and the volume of the system, respectively. 
With the help of diagrammatic techniques, it is possible to show that \cite{hansen2013theory, santos2016concise}
\begin{equation}\label{eq:bridge}
    g(r) = \exp(-\beta u(r) + h(r) - c(r) + b(r)),
\end{equation}
in which $\beta u(r)$ is the interaction potential in units of thermal energy ($\beta = 1/k_BT$), and $b(r)$ is the so-called bridge function, which is the sum of all `elementary' diagrams. Because no exact expression for the bridge function is known, it is necessary to introduce approximate closure relations, which either express the bridge function in terms of $c(r)$, $h(r)$, and $\beta u(r)$ or approximate it in some other fashion. Together with a closure relation, a system of three equations is formed that can be solved self-consistently. Independent of the form of the bridge function chosen, Eqs.~\eqref{eq:oz} and \eqref{eq:bridge} already mathematically constrain the possible solutions to the equations. In particular, \citet{kast2012communication} found that $b(r) \leq c(r) + \beta u(r)$ must hold for all $r$.
While several exact solutions exist for simple model systems and closures (the most famous being perhaps of \citet{wertheim1964analytic} and \citet{thiele1963equation}), the equations typically must be solved numerically. 

For mixtures of different particle species, Eq.~\eqref{eq:oz} becomes matrix-valued, and Eq.~\eqref{eq:bridge} is applied for each species pair separately. In this work, however, we focus on the simpler case of single-component systems to keep the results tractable. (We have ensured, however, that the published code \cite{OZjl} also works for mixtures.)

In the case of particles with discontinuous interaction potentials, such as hard spheres, discontinuities arise also in the direct and total correlation functions, $c(r)$ and $h(r)$. Their difference $\gamma(r) = h(r) - c(r)$, called the indirect correlation function,  remains continuous however, which is useful both for numerical and theoretical reasons \cite{hansen2013theory}. Similarly, the cavity distribution function $y(r) = \exp(\beta u(r))g(r)$  and the bridge function $b(r)$ can be shown to be continuous. These observations invite us to reformulate Eq.~\eqref{eq:bridge} as
\begin{align}
    \ln y(r) = \gamma(r) + b(r).
\end{align}
This form is useful for theoretical purposes as well as for determining the bridge function from simulations.


The solution of Eqs.~\eqref{eq:oz} and \eqref{eq:bridge} not only gives access to structural properties but also thermodynamic ones. By virtue of the virial equation, the pressure $p$ can be computed from the pair structure as  \cite{santos2016concise}

\begin{equation}\label{eq:virial}
    \frac{\beta p}{\rho} = 1 - \frac{\rho}{2d}\int \mathrm{d} \textbf{r} \beta u'(r) r g(r)
\end{equation}
in $d$ dimensions, where $u'(r)$ is the derivative of the pair potential $u(r)$. The total potential energy $U$ is equal to the sum of the pair interactions, which can be rewritten as
\begin{equation}
    \frac{U}{N} = \frac{\rho}{2} \int \mathrm{d} \textbf{r} u(r) g(r)
\end{equation}
and for the isothermal compressibility $\chi_T$ it can be shown that
\begin{equation}\label{eq:compressibility}
    \rho k_BT\chi_T =  k_BT\left(\frac{\partial \rho}{\partial p}\right)_\beta = 1+\rho\int \mathrm{d}\textbf{r}h(r).
\end{equation}


Equation \eqref{eq:compressibility} suggests that there is a consistency requirement between the compressibility Eq.~\eqref{eq:compressibility}, and that obtained by differentiating the pressure obtained from the virial route, Eq.~\eqref{eq:virial}:
\begin{align}
    \frac{\partial}{\partial \rho} \left[\rho - \frac{\rho^2}{2d}\int \mathrm{d} \textbf{r} \beta u'(r) r g(r; \rho)\right]&=\frac{1}{1+\rho\int \mathrm{d}\textbf{r}h(r)} \nonumber\\ &= 1-\rho\int \mathrm{d}\textbf{r}c(r) \label{eq:consistency} .
\end{align}
Similarly, since 
\begin{equation}\label{eq:consistency2}
    \left(\frac{\partial U}{\partial V}\right)_{\beta} = -\left(\frac{\partial \beta p}{\partial \beta} \right)_{V},
\end{equation}
we have a second consistency relation:
\begin{align}
    0 = \int \mathrm{d}\textbf{r}& \left\{u(r)\left(g(r) + \rho \left(\frac{\partial g(r)}{\partial \rho}\right)_T\right) \right.\\&+\left. \frac{u'(r) r}{d}\left[g(r) - k_BT \left(\frac{\partial g(r)}{\partial k_BT}\right)_\rho\right]\right\}.\nonumber
\end{align}
In real systems, these are always satisfied, but the same does not hold in general for the solution of Eqs.~\eqref{eq:oz} and \eqref{eq:bridge} when an approximate closure is used. Many closures have been developed that by design satisfy one or both of the thermodynamic consistency relations \eqref{eq:consistency} and \eqref{eq:consistency2}. While this typically improves the accuracy of the theory, it does not guarantee the qualitative correctness of $g(r)$. Moreover, even closures that incorporate computer simulation data of thermodynamic properties are not guaranteed to predict accurate correlation functions.

It is important to point out that many closures in the literature adhere to the unique functionality conjecture, which asserts that $b(r)$ is a unique local function of $\gamma(r)$, that is, $b(r) = b(\gamma(r))$ \cite{lee1992chemical}. While approximately true, this is known to be inaccurate for certain ranges of $r$ at high densities \cite{llano1994bridge, llano1992bridge, kolafa2002bridge, francova2010accurate}. We will revisit this later. 

\subsection{Closure relations}

In Table \ref{tab:closures} we list in chronological order the closure relations studied in this work. For conciseness, we shall not discuss here the motivations behind the closures but instead refer to the excellent reviews by \citet{bomont2008recent, solana2013perturbation}, the historical account of \citet{rowlinson1965equation} for the earlier closures, and the references in Table \ref{tab:closures}. 

While we consider a large number of closures in this work, it has been inevitable to exclude many others. For example, we have not included historical approximations \cite{rushbrooke1953theory, salpeter1958mayer}, second order theories \cite{attard1989spherically, lomba2000inhomogeneous, kim2002three, brader2008structural, lee2011constructing} (because of their numerical complexity), closures that are very close in nature to others considered in this work \cite{rowlinson1965self, hutchinson1972thermodynamically, hall1980thermodynamically, rosenfeld1979theory}, closures developed for model systems not considered here \cite{lebowitz1966mean, blum1998analytical, amokrane2005structure}, and machine-learned closures \cite{goodall2021data, bedolla2021using, carvalho2020indirect}. In addition, fundamental measure theory results can also be used as a closure for the Ornstein-Zernike equation \cite{rosenfeld1989free, roth2010fundamental}, but we exclude those as well because they do not directly apply to systems with generic interaction potentials.

\begin{table*}[t]
    \centering
    \renewcommand{\arraystretch}{1.6}
    \caption{Closures used in this work enumerated chronologically.  Many closure relations listed have free parameters. Unless otherwise stated, for those with a single free parameter $\alpha$, Eq.~\eqref{eq:consistency} is used to obtain it. For the ERY closure, Eq.~\eqref{eq:consistency} is used in conjunction with Eq.~\eqref{eq:consistency2}. }
    \label{tab:closures}
    \begin{tabular}{|c|l|c|c|c|}
    \hline
        & \textbf{Closure name}  & \textbf{Abbreviation} & \textbf{Definition, $b(r) = $} & \textbf{Reference}   \\\hline
        1. & Percus-Yevick  & PY &  $\ln(1+\gamma(r)) - \gamma(r)$ & \citen{percus1958analysis}\\\hline
        2. & Hypernetted chain  & HNC &  $0$ & \citen{van1959new,meeron1960nodal,morita1960new, rushbrooke1960hyper, verlet1960theory}\\\hline
        3. & Modified hypernetted chain$^a$& MHNC&  $b_\mathrm{HS}(r)$ & \citen{rosenfeld1979theory}\\\hline
        4. & Verlet  & V &  $-\frac{1}{2}\gamma^2(r)/{(1+4\gamma(r)/5)}$ & \citen{verlet1980integral}\\\hline
        5. & Modified Verlet$^{b}$  & MV &  $-\frac{1}{2}\gamma^2(r)/{(1+\alpha\gamma(r))}$ & \citen{verlet1980integral}\\\hline
        6. & Martynov-Sarkisov  & MS &  $\sqrt{1+2\gamma(r)} - \gamma(r) - 1$ & \citen{martynov1983exact} \\\hline
        7. & Rogers-Young  & RY &  $\ln\left(1+\frac{\exp(f(r)\gamma(r))-1}{f(r)}\right)-\gamma(r)$; $f(r) = 1-\exp(-\alpha r)$ & \citen{rogers1984new} \\\hline
        8. & Ballone-Pastore-Galli-Gazzillo$^b$  & BPGG &  $\left(1+\alpha\gamma(r)\right)^{1/\alpha} - \gamma(r) - 1$ & \citen{ballone1986additive}\\\hline
        9. & Lee$^{de}$ & L &  $-\frac{1}{2}\alpha_1 \tilde\gamma(r)^2\left[1-(\alpha_2 \alpha_3 \tilde\gamma(r))/(1 + \alpha_3\tilde\gamma(r))\right]$ & \citen{lee1995accurate}\\\hline
        10. & Duh-Haymet$^{b}$& DH &  $-\frac{1}{2}\gamma^2(r)\left[1+\gamma(r)\left(5\gamma(r)+11\right)/\left(7\gamma(r)+9\right)\right]^{-1}$ & \citen{duh1995integral, duh1996integral}\\\hline
        11. & Charpentier-Jackse  & CJ &  $\left(\sqrt{1+4\alpha \gamma(r)}-1-2\alpha\gamma(r)\right)/2\alpha$ & \citen{charpentier2001exact}\\\hline
        12. & Choudhury-Ghosh$^{bf}$  & CG &  $-\frac{1}{2}\gamma^2(r)/(1+\zeta(\rho)\gamma(r))$ & \citen{choudhury2002integral}\\\hline
        13. & Bomont-Bretonnet  & BB &  $\sqrt{1+2\gamma(r) + \alpha\gamma^2(r)}-1-\gamma(r)$ & \citen{bomont2003self}\\\hline
        14. & Extended Rogers-Young & ERY & $\ln\left(1+\phi(r) + \alpha_2 \phi(r)^2\right)-\gamma(r)$; $\phi(r) = \frac{e^{(1-\exp(-\alpha_1 r))\gamma(r)}-1}{1-\exp(-\alpha_1 r)}$ & \citen{carbajal2008thermodynamically}\\\hline
        15. & Carbajal-Tinoko$^{c}$  & CT &  $e(r;\alpha)\left[(2-\ln y(r))y(r)-2-\ln y(r)\right]/\left(y(r)-1\right) $ & \citen{carbajal2022local} \\\hline
    \end{tabular}\\
    \footnotesize{$^a$ The function $b_\mathrm{HS}(r)$ is the bridge function of a hard-sphere system that is parameterized from simulation data. We use the one of \citet{malijevsky1987bridge}. The volume fraction of the reference system can be chosen freely, and can thus be chosen to fulfill a consistency relation.  For the hard-sphere case specifically, we do not treat the volume fraction occurring in the MHNC closure as a free parameter but set it equal to the actual volume fraction of the system. }
    \footnotesize{$^b$ These closures additionally have $b(r) = -\gamma^2(r)/2$ if $\gamma(r)<0$.}
    \footnotesize{$^c$ The function $e(r;\alpha)=3\exp(\alpha r)$ for $\alpha <0$ and $e(r;\alpha) = 3+\alpha$ otherwise.}
    \footnotesize{$^d$ The parameters $\alpha_i$ are to be determined by thermodynamic consistency relations and zero-separation theorems, see \citet{lee1995accurate} and the references therein. In particular, the coefficients are chosen such that the Carnahan-Starling hard-sphere results for the zero-separation relations are satisfied. We were unable to reproduce the reported results of \citet{lee1995accurate}, and also to satisfy all conditions robustly. Therefore, instead, we determine $\alpha_i$ by minimizing the sum of squares of the three conditions \cite{fernaud2000self}.}
    \footnotesize{$^e$ For this closure, $\tilde\gamma(r) = \gamma(r)+\rho \sigma^d\left(\exp(-\beta u(r))-1\right)/2$.}\\
    \footnotesize{$^f$ The function $\zeta(\rho)=1.0175-0.275\rho\sigma^d$ is an empirically determined relation for the Lennard-Jones fluid in $d=3$, which we employ for all systems in this work.}

\end{table*}

Before we discuss our methods and results, let us briefly point out some notational conventions and considerations regarding closure relations.
Specifically, when the interaction potential has an attractive tail, the bridge function of the corresponding system is more suitably approximated by a functional of the so-called renormalized indirect correlation function $\gamma^*(r)=\gamma(r) - \beta u_{\mathrm{LR}}(r)$, where $u_{\mathrm{LR}}(r)$ is the attractive ``long-ranged" part of the potential \cite{madden1980mean, zerah1986self}. This is the reason that many of the closures in Table \ref{tab:closures} have been adapted to be a function of $\gamma^*(r)$ instead of $\gamma(r)$. Such closures we indicate with an asterisk in the abbreviation. For example, the well-known soft-core mean-spherical approximation (SMSA) \cite{chihara1973integral, madden1980mean} is denoted as PY$^*$ (since it is equivalent to the Percus-Yevick approximation when $\gamma\leftrightarrow\gamma^*$). Similarly, the hypernetted-chain-SMSA closure, devised by \citet{zerah1986self}, is denoted as RY$^*$ since it becomes equivalent to the Rogers-Young (RY) approximation. 

How the interaction potential is split into a short- and long-ranged part can be chosen freely. Often, the splitting famously introduced by \citet{weeks1971role} is employed, which states that $u(r)=u_\mathrm{SR}(r) + u_\mathrm{LR}(r)$, where
\begin{equation}
    u_\mathrm{SR}(r) = 
    \left\{\begin{array}{lr}
        u(r) - u(r^*), \qquad &\,\,\,\,\,r< r^*,\\
        0,  \qquad &\,\,\,\,\,r>r^*,
    \end{array}\right.
\end{equation}
and
\begin{equation}
    u_\mathrm{LR}(r) = 
    \left\{\begin{array}{lr}
        u(r^*), \qquad &\,\,\,\,\,r< r^*,\\
        u(r),  \qquad &\,\,\,\,\,r>r^*,
    \end{array}\right.
\end{equation}
in which $r^*$ is typically chosen to coincide with the minimum of the potential. 
Different splitting schemes have also been proposed. For example, \citet{duh1995integral} suggested the use of 
\begin{equation}
    u_\mathrm{LR}(r) = -4\epsilon\left(\frac{\sigma}{r}\right)^6\exp\left[\frac{-\epsilon}{\rho}\left(\frac{\sigma}{r}\right)^6\right]
\end{equation}
specifically for the Lennard-Jones potential (where $\epsilon$ and $\sigma$ are the canonical Lennard-Jones parameters). Other procedures have also been proposed \cite{barker1967perturbation, mcquarrie1966high} We will briefly address the splitting procedures in Section \ref{sec:LJ}. 

\section{Methods}

\subsection{The Ornstein-Zernike equation}\label{sec:numoz}
Throughout this work, we solve the Ornstein-Zernike equation, together with its closure by the method of \citet{ng1974hypernetted}. Essentially, the method consists of the following steps:
\begin{enumerate}
    \item Start with an initial guess $\gamma^{(0)}(r)$;
    \item Use the chosen closure relation to compute $c^{(0)}(r) = e^{-\beta u(r) + \gamma^{(0)}(r) + b^{(0)}(r)}-\gamma^{(0)}(r)-1$;
    \item Compute its Fourier transform $\hat{c}^{(0)}(k)$;
    \item Compute Fourier transform of the indirect correlation function with the Ornstein-Zernike relation $\hat{\gamma}^{(1)}(k) = \rho \hat{c}^{(0)}(k)^2/(1 - \rho\hat{c}^{(0)}(k))$ ;
    \item Transform back to find a better guess $\gamma^{(1)}(r)$ for the indirect correlation function;
    \item Repeat steps 2--5 until convergence.
\end{enumerate}
To speed up the convergence, \citet{ng1974hypernetted} proposed to use an optimal mixing rule taking into account the current and a number $\ell$ earlier iterations of $c(r)$ at step 2. We use their iteration procedure with $\ell=4$ throughout this work. 

The radially symmetric Fourier transforms in $d$ dimensions are given by
\begin{equation}
  \hat{f}(k) = (2\pi)^{d/2} k ^{1-d/2}\int_0^\infty \mathrm{d}r r^{d/2}J_{d/2-1}(kr)f(r)
\end{equation}
and
\begin{equation}
f(r) = (2\pi)^{-d/2} r ^{1-d/2}\int_0^\infty \mathrm{d}k k^{d/2}J_{d/2-1}(kr)\hat{f}(k),  
\end{equation}
where $J_{p}(x)$ is the $p$th order Bessel function of the first kind, and $k$ is the radial wave number. For the specific case that $d=3$, the transforms reduce to
\begin{equation}
\hat{f}(k) = \frac{4\pi}{k}\int_0^\infty \mathrm{d}r r \sin(kr)f(r),
\end{equation}
 and
\begin{equation}
f(r) = \frac{1}{2\pi^2r}\int_0^\infty \mathrm{d}k k\sin(kr)\hat{f}(k).
\end{equation}
The latter is implemented using the type-4 fast discrete sine transform provided by the FFTW software package \cite{frigo1998fftw}. This results in an error convergence rate of the Ornstein-Zernike iteration of second order in the number of discretization points. In the $d$-dimensional case, a scheme with first-order accuracy is used for the Fourier-Bessel transforms. Because of the iterative nature of the scheme, orthogonality must be ensured. To ensure convergence of the Fourier transform we discretized the correlation functions on $M=10^4$ points with separation $\Delta r=0.001$ in $d=3$ and $\Delta r = 0.003$ for $d=2$. The larger step size $\Delta r$ in two dimensions is necessary since correlation functions in that case are typically of longer range.

\subsection{Monte Carlo simulations}\label{sec:numMC}\label{sec:bridge}

To test the predictions of the integral equation theories, we perform Metropolis Monte Carlo simulations in the canonical \textit{NVT} ensemble. The number of particles in the simulation is given by $N$~=~30\,000 for the hard-sphere system and by $N$~=~10\,000 for all other systems. We use a cut-off of the potential at $r_c = 5\sigma$ for all potentials except for hard spheres. Thus, the pair potential used in the simulations is equal to 
\begin{equation}
    u_\mathrm{MC}(r) = 
    \left\{\begin{array}{lr}
        u(r) - u(r_c), \qquad &\,\,\,\,\,r< r_c,\\
        0,  \qquad &\,\,\,\,\,r>r_c.
    \end{array}\right.
\end{equation}
After equilibration, the liquids are simulated for at least $3\times 10^6$ attempted moves for every particle, during which the trajectories are saved every $10^3$ steps, from which the correlation functions are computed. We use periodic boundary conditions to minimize finite-size effects.

The bridge function is obtained by 
\begin{equation}
    b(r) = \ln y(r) - \gamma(r)
\end{equation}
where $\gamma(r)$ is obtained by direct simulation of the radial distribution function $g(r)$. Specifically, once $g(r)$ is obtained, $g(r)-1$ is numerically Fourier transformed and the indirect correlation function is obtained by use of the Ornstein-Zernike relation $\hat{\gamma}(k) = \rho \hat{h}(k)^2/(1 + \rho\hat{h}(k))$, which is then transformed back. The Fourier transforms used are more accurate quadratures than those described in Section \ref{sec:numoz}. We present them in Appendix \ref{app:quad}. For additional accuracy, we apply a finite size correction as outlined in \citet{kolafa2002bridge}, who argued that the finite-size errors of $g(r)$ are much smaller in the grand canonical ensemble than in the canonical one. Therefore, they show, one may correct the radial distribution function
\begin{align}
    g(r) &\approx g_{\mu VT}(r) \\&= g_{NVT}(r)+\frac{k_BT}{2N}\left(\frac{\partial \rho}{\partial p}\right)_T \frac{\partial^2 \rho^2 g(r)}{\partial \rho^2} + \mathcal{O}(N^{-2})\nonumber.
\end{align}
This corrects for the spurious $g(r\to\infty)= 1-1/N$ limit obtained in the $NVT$-ensemble. We evaluate both the inverse compressibility $\left(\frac{\partial \rho}{\partial p}\right)_T$ and the second derivative of $g(r)$ with respect to the density by the solution of the Ornstein-Zernike equation with the HNC closure, which is sufficiently accurate here as it concerns only a $1/N$ correction of the radial distribution function. For more details on this procedure, we additionally refer to the work by \citet{castello2022bridge}.

\begin{figure*}[ht]
    \centering
    \includegraphics[width=\linewidth]{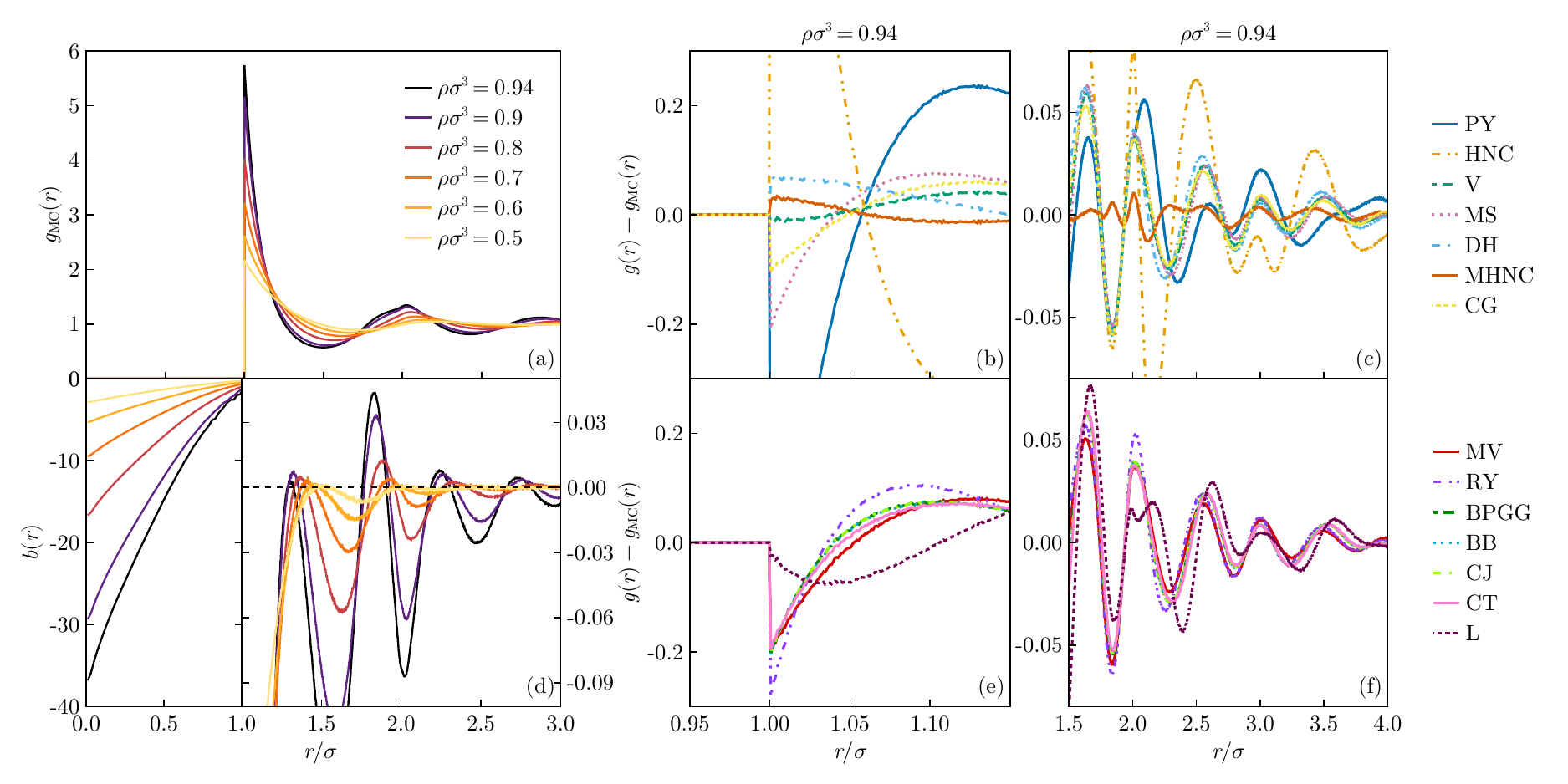}
    \caption{Hard sphere system: accuracy of predicted radial distribution function $g(r)$. In panel (a), we show $g(r)$ for various densities obtained from Monte Carlo simulations. Panels (b--c, e--f) show the difference of the Monte Carlo data with that obtained from a self-consistent solution of the Ornstein-Zernike equation with the various closures. In particular, the bottom two panels focus on thermodynamically consistent closures, whereas the top two do not have a free parameter. At $\rho\sigma^3 = 0.94$, the deviation of PY at contact $g_\mathrm{PY}(\sigma)-g_\mathrm{MC}(\sigma) = -0.921$, whereas for HNC it is $g_\mathrm{HNC}(\sigma)-g_\mathrm{MC}(\sigma) = 1.966$, where $\sigma$ is the sphere diameter. Panel (d) shows the bridge function $b(r)$ for different densities. A different vertical scale is used for the regimes $r<\sigma$ and $r>\sigma$. The ERY closure is not applied because it relies on Eq.~\eqref{eq:consistency2}, which is always satisfied for the hard sphere system.}
    \label{fig:HS-gr}
\end{figure*}

For the cavity distribution function $y(r) = g(r) \exp(\beta u)$, two complementary methods are employed. In the region where $\beta u\gg 1$, we have $g(r)\ll 1$, and thus the simulated $g(r)$ data are not sufficiently accurate to obtain a satisfactory $y(r)$. This is even clearer in the hard-sphere case, where this method fails entirely. To circumvent this issue, $y(r)$ is simulated directly with a test-particle method described in \citet{henderson1983test}, as well as in \citet{llano1992bridge}. The basic idea is that the cavity distribution function can be obtained by creating a cavity and measuring the energy it costs to insert a particle some distance $r$ away from that cavity. Since this procedure is already central to standard Monte Carlo moves, where one measures the energy cost $\Delta E = E_2 - E_1$ of moving a particle from position 1 to position 2 (creating a cavity at 1), this measurement can be performed cheaply. The cavity distribution function is now given by
\begin{equation}
    y(r) = \langle \exp (\beta \mu_\mathrm{ex}-\beta E_2) \rangle,
\end{equation}
where $\mu_\mathrm{ex}$ is the excess chemical potential, which we have determined by Widom's test particle method \cite{widom1963some, frenkel2023understanding}, and the angular brackets denote ensemble averaging. 

In the case where $\beta u(r) \not\gg 1,$ and $g(r)\approx 1$, we can use the definition of the cavity distribution function $y(r)=g(r)\exp(\beta u(r))$ directly. Where applicable, the latter provides much more accurate bridge functions.

\section{Results}
In this section, we test the various closures of the Ornstein-Zernike relation against Monte Carlo simulation data. We consider systems with the following pair potentials:
\begin{itemize}
    \item Hard spheres; $u(r<\sigma)=\infty$, $u(r>\sigma)=0$
    \item Inverse power-law; $u(r) = \epsilon(\sigma/r)^n$
    \item Lennard Jones; $u(r) = 4\epsilon\left[(\sigma/r)^{12}-(\sigma/r)^6\right]$
    \item Gaussian Core; $u(r) = \epsilon\exp(-r^2/\sigma^2)$
\end{itemize}
in which $\epsilon$ and $\sigma$ are parameters denoting the strength of the potential and size of the particles, respectively. As mentioned in the methods, each of the potentials is cut off and shifted at $r_c = 5\sigma$ for computational efficiency.
These systems were chosen because they encompass different classes of interaction potentials, including both attractive and repulsive as well as impenetrable and penetrable ones. We consider the two-dimensional and three-dimensional systems for each interaction potential. For the eight different systems, we have performed Monte Carlo simulations of the corresponding system at different state points. From the trajectories, we have computed the radial distribution functions and bridge functions, which we compare to the predictions of the different closures. We focus on comparing correlation functions directly instead of using derived thermodynamic properties because good predictions of the former do not necessarily imply the same for the latter as we show in Section \ref{sec:GCM}.

\subsection{Hard spheres (3D)}

\begin{figure}[t]
    \centering
    \includegraphics[width=\linewidth]{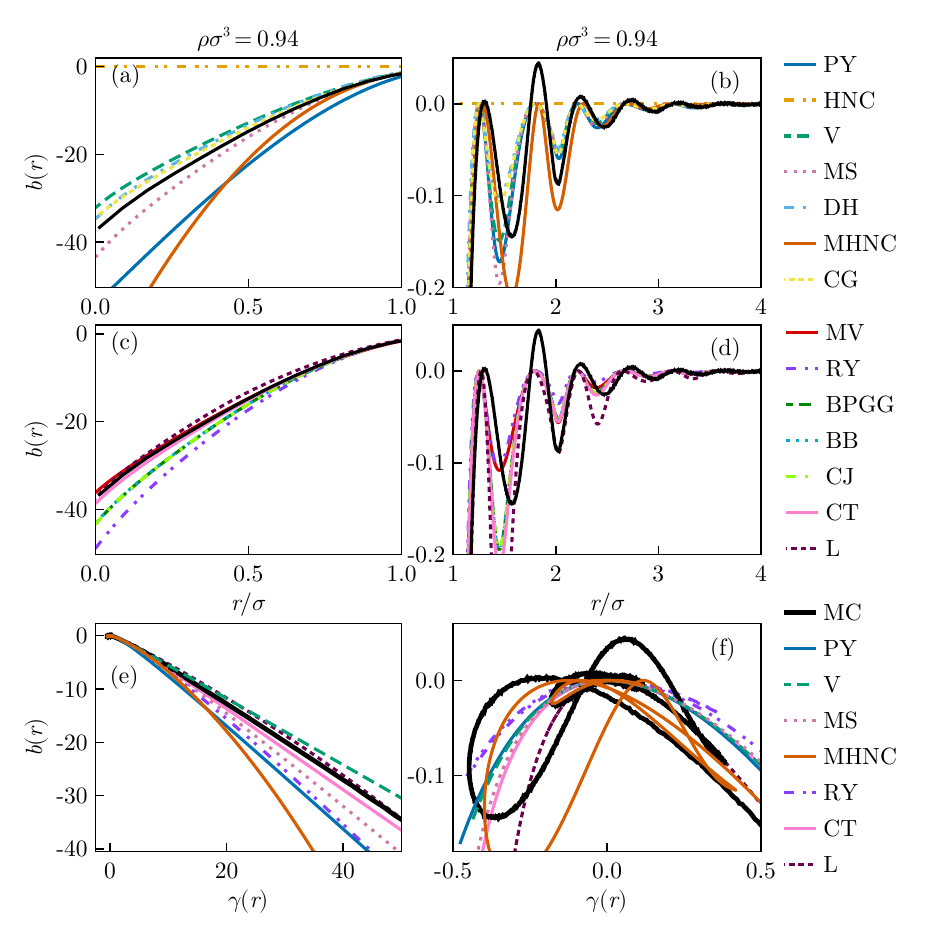}
    \caption{Hard sphere system: comparing the bridge function obtained from the simulation data and self-consistent solution of the Ornstein-Zernike equation with various closure relations at density $\rho\sigma^3=0.94$. The solid black line indicates the results from the Monte Carlo simulations. In panels (a--b), we show the comparison for closures without a free parameter, for $r<\sigma$ (a) and for $r>\sigma$ (b). For closures with a free parameter that fixes thermodynamic consistency, we show the same comparison in (c) and (d) respectively. Panels (e--f) show Duh-Haymet plots of $b(r)$ as a function of $\gamma(r)$. Here, panel (f) is a magnification of (e).}
    \label{fig:HS-br}
\end{figure}

The hard-sphere system is perhaps the most studied in relation to the Ornstein-Zernike equation, and most, if not all, of what we present in this section is a replication of what is presented elsewhere. We have chosen, however, not to omit it, because we feel there is value in an encompassing overview of the accuracy of the different closures for this system as well. The comparison with results published previously also serves as a good validity test of our methods. 

The radial distribution and bridge functions of this system are shown for reference in Fig.~\ref{fig:HS-gr}a. Our data of the bridge function very accurately agrees with that of \citet{kolafa2002bridge}. In Fig.~\ref{fig:HS-gr}(b--e), we compare at a high density ($\rho\sigma^3=0.94$, just below the melting point) the radial distribution function obtained from Monte Carlo simulations, with that obtained from the solution of Eqs.~\eqref{eq:oz} and \eqref{eq:bridge}, together with several closures. Specifically, in Fig.~\ref{fig:HS-gr}(b--c), we consider closures that do not include a free parameter, and in Fig.~\ref{fig:HS-gr}(d--e) those that do. The free parameters were obtained by the pressure consistency route, Eq.~\eqref{eq:consistency}, and, specifically in the case of the Lee (L) closure, we also used the Carnahan Starling parametrization for the zero-separation theorems as described in \citet{lee1995accurate}. We have checked that our numerically obtained $g(r)$ for the Percus-Yevick closure agrees with that obtained from the exact expression, that the pressures from the Rogers-Young closure agree accurately with those reported in \citet{rogers1984new}, and we can reproduce all decimal places of the values of the thermodynamic properties and the self-consistency parameters $\alpha$ reported in table I of \citet{carbajal2008thermodynamically}. However, we cannot accurately reproduce the values of $b(0)$ reported in Table I of \citet{lee1995accurate}, which we attribute to a difference in a numerical integration scheme (noting also that the values reported there do not accurately agree with the exactly known values for PY).

Our results in Fig.~\ref{fig:HS-gr} show the predictions of the closures for the radial distribution functions. Panels (b,e) and (c,f) show the errors of the predicted radial distribution function in the region near the first peak and of the subsequent peaks respectively. Focusing our attention to panel (b), it is clear that the well-known HNC and PY approximations perform the worst out of all closures considered, especially near the peak of $g(r)$. PY famously performs better than HNC \cite{hansen2013theory}, the former of which can be improved further by the \textit{ad hoc} correction of \citet{verlet1972equilibrium}.  We can see that among the other closures without free parameters, the Verlet closure reproduces the contact value the best, closely followed by the MHNC closure, whose good performance is of course not surprising because the closure is essentially a parametrization of the bridge function obtained from computer simulations. As such, we are effectively validating the work of \citet{malijevsky1987bridge}, whose parameterization we use.

Our results of panel (e) show that we find no large qualitative differences between performances of the thermodynamically consistent closures, except for that of \citet{lee1995accurate}, which gives a better agreement of the pair correlation function at contact $g(\sigma)$ (and concomitantly the pressure) because the closure implicitly incorporates the Carnahan-Starling equation of state. Panels (c) and (f) show that for larger distances the MHNC closure qualitatively and quantitatively outperforms the other closures. In this regime, the errors that many closure relations make are very similar, the reason for which we discuss below.

Let us focus our attention now on the accuracy of the predicted bridge function. In Fig.~\ref{fig:HS-br}(a--d), we show the function obtained from the integral equation theories together with simulation data. Again, we show separately the thermodynamically consistent closures and those that are not. In the figure, the left panels represent distances where the spheres overlap ($r<\sigma$), while the right panels depict states where there is no overlap ($r>\sigma$). Unsurprisingly, the data show that, overall, those closures that produce an accurate $g(r)$ also represent $b(r)$ accurately. A notable exception is the MHNC closure, whose parameterization was performed only over the region $r>\sigma$, resulting in its failure for distances below contact. It is important to note that while good predictions for the bridge function in the $r<\sigma$ regime are not very important for accurate $g(r)$'s, they are important for thermodynamic properties \cite{lee1995accurate}. 

In the $r>\sigma$ regime, all closures assert that $b(r)$ remains negative, while in reality, this is not the case \cite{llano1994bridge, llano1992bridge, kolafa2002bridge, francova2010accurate}. In panels (e--f) we show the bridge function as a function of the indirect correlation function. This is often referred to as a Duh-Haymet plot \cite{duh1995integral}. It shows that in the regime where $r<\sigma$ (where both $\gamma(r)$ and $-b(r)$ are large), the bridge function $b(r)$ can indeed be written as a well-behaved function of $\gamma(r)$. However, in the region $r>\sigma$, which is more important for the accuracy of $g(r)$, this is unfortunately not the case. Here, many closures that give $b=b(\gamma(r))$ fail because $\gamma(r)$ cannot be mapped uniquely to $b(r)$ \cite{llano1994bridge, llano1992bridge, kolafa2002bridge, francova2010accurate}. Due to their similar asymptotic series at small $\gamma(r)$ (typically $b(r)\approx -\gamma(r)^2/2$)\cite{bomont2003self}, many closures break down in the same way in this regime, causing the behavior shown in Fig.~\ref{fig:HS-gr}(c,e). Exceptions are the closures MHNC, RY, CT, and L, which do not predict that $b(r)$ is a pure function of $\gamma(r)$, but have an explicit dependence on $r$: $b=b(\gamma(r), r)$. However, in the latter three cases, the dependence on $r$ is not sufficiently strong to avoid this issue. For MHNC, its parameterization does not depend on $\gamma(r)$ at all, and it therefore does not experience the same limitation. Our findings agree well with those reported in \citet{lee1992chemical}, who report a similar comparison with four closures at $\rho\sigma^3=0.8$. 

\subsection{Hard disks (2D)}

\begin{figure}[t]
    \centering
    \includegraphics[width=\linewidth]{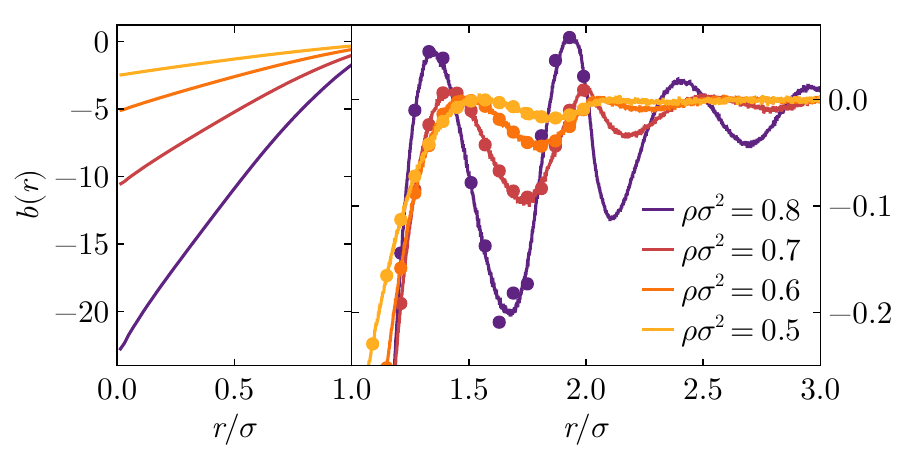}
    \caption{Hard disk system: bridge function at different densities. The left panel shows the overlap region ($r<\sigma$) and the right panel shows the region $r>\sigma$. In the right panel, the markers show data obtained by direct evaluation of the cavity distribution function, and the lines are obtained through the radial distribution function. Both methods are described in section \ref{sec:bridge}.}
    \label{fig:bridge_HD}
\end{figure}

The hard disk system in two dimensions is the equivalent of the hard-sphere system in three dimensions. In contrast to the three-dimensional case, in two dimensions the system features a discontinuous liquid-hexatic transition around $\rho\sigma^2=0.89$ and subsequently a continuous hexatic-solid transition \cite{bernard2011two, engel2013hard, kapfer20142d}. Approaching the liquid-hexatic transition from the liquid side, the orientational correlation length grows drastically, which has necessitated increasingly large simulations to properly establish and quantify the emergence of the hexatic phase. In this work, however, we remain on the liquid side of the transition and consider densities up to (only) $\rho\sigma^2=0.8$. In this regime, our simulations with $N=10\,000$ particles remain sufficiently large to observe the decay of spatial correlation functions.

\begin{figure}[t]
    \centering
    \includegraphics[width=\linewidth]{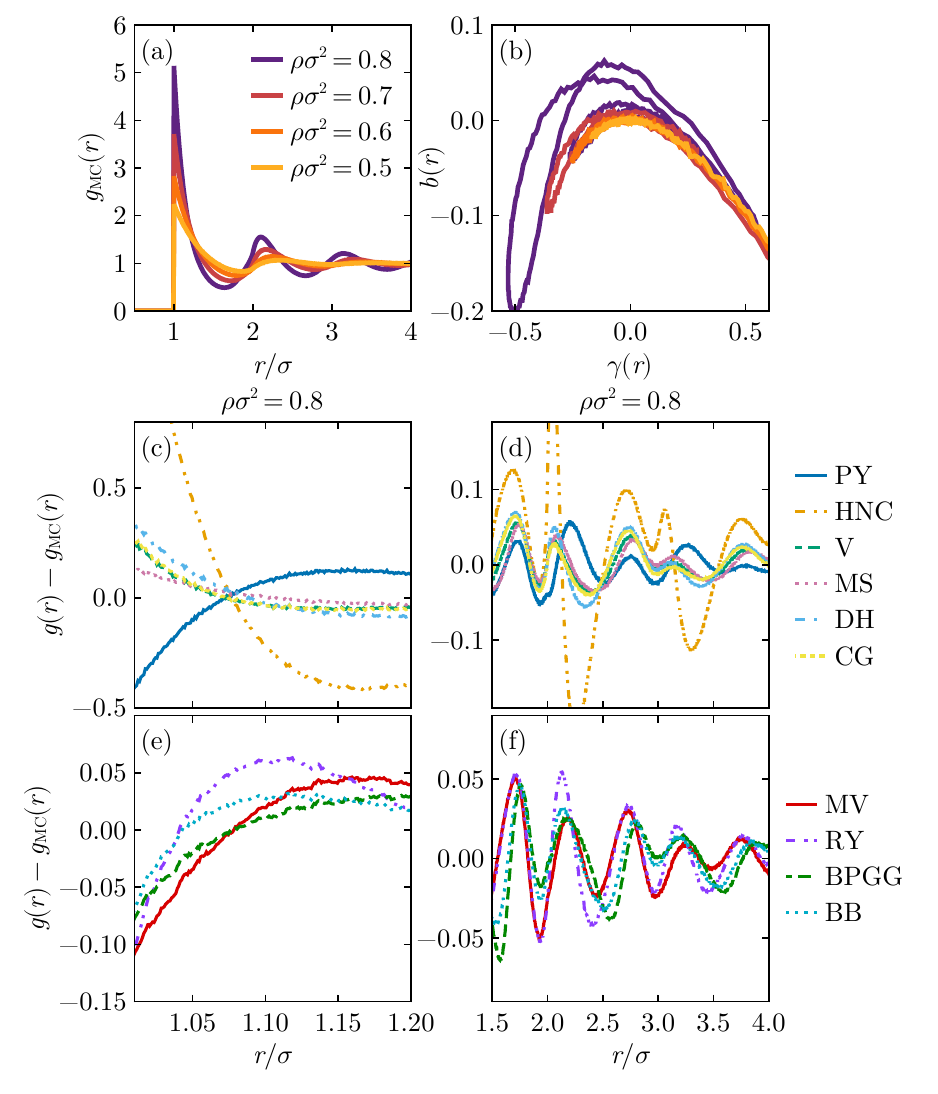}
    \caption{Hard disk system: comparison of the accuracy of the predicted radial distribution function $g(r)$. In panel (a), we show $g(r)$ for various densities obtained from Monte Carlo simulations. Panel (b) shows the DH plot of $b(r)$ against $\gamma(r)$ for different densities. Panels (c--f) show the difference of the Monte Carlo data with that obtained from a self-consistent solution of the Ornstein-Zernike equation with the various closures. In particular, the bottom two panels focus on thermodynamically consistent closures, whereas the middle two do not have a free parameter. 
     Only closures that converged to physical results are shown.}
    \label{fig:Hard_disk_closures}
\end{figure}

To the best of our knowledge, there is no data available in the literature for the bridge function of the hard-disk liquid (or any 2D fluid for that matter). Perhaps this is because the two-dimensional radial Fourier transform is more difficult to perform accurately than its three-dimensional counterpart. In the Appendix, we present a simple generalized midpoint rule that we developed for this purpose. We show the resulting bridge function in Fig.~\ref{fig:bridge_HD} for different densities. In the right panel, we use symbols and lines to show the bridge function obtained by the two separate methods to confirm their accuracy.

Comparing the bridge function of the hard-disk system to that of the hard-sphere system (inset of Fig.~\ref{fig:HS-gr}a) we see that the functional forms seem to be very similar. The qualitative behavior in the overlap region is identical, and in the region $r>\sigma$, both oscillate with similar frequency and phase (slightly depending on density) and both feature non-negative peaks. The most notable difference is that relative to the zero-separation values $b(0)$, the first peaks are much higher in the two-dimensional system. Additionally, in the three-dimensional system, the first peak is consistently much lower than the second one. This difference in peak height persists in the two-dimensional case, but it is much less pronounced.

In Fig.~\ref{fig:Hard_disk_closures}, we compare the accuracy of various closures for the high-density hard disk system fluid. A number of closures have been excluded from the figure, due to convergence issues at this state point. Overall, we conclude that most trends that are visible in the performances of the closure theories for the hard sphere case are also present in its two-dimensional equivalent. In particular, again PY outperforms HNC, the former underestimating, the latter overestimating $g(\sigma)$. In turn, they are outperformed by all other closures tested (excluding those that did not converge). Among the thermodynamically consistent integral equation theories, all underestimate the $g(\sigma)$ by around two percent, which is better than any of the non-consistent ones. For $r>3\sigma/2$, the performance of the consistent closures is comparable to that of the non-consistent ones (excluding HNC).

\subsection{Inverse power law (3D)}

\begin{figure}[t]
    \centering
    \includegraphics[width=\linewidth]{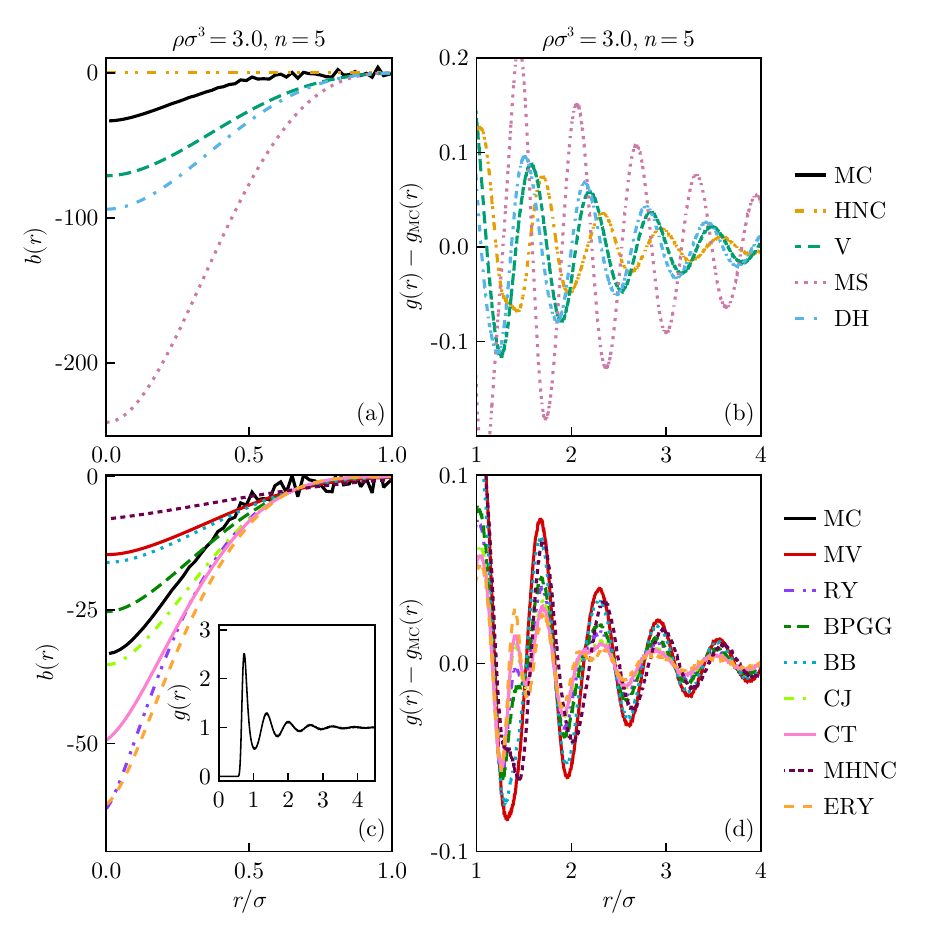}
    \caption{3D inverse power-law system: bridge function $b(r)$ (a, c) and radial distribution function $g(r)$ (b, d)  compared to those obtained from Monte Carlo simulations. The top panels show thermodynamically inconsistent closures, and the bottom panels show the consistent ones. Note that the vertical scales of (c, d) are smaller compared to (a, b). The PY and CG closures converged to non-physical results and are therefore not shown. The inset of panel (c) shows the radial distribution function at this state point.}
    \label{fig:IPL}
\end{figure}

The inverse power law potential $u(r)=\epsilon \left(r/\sigma\right)^{-n}$ can represent a wide range of particles that have repulsive interactions, but that are softer than hard spheres \cite{prestipino2005phase, hoover1971thermodynamic, hoover1972statistical, vovcadlo2002ab, scalliet2022thirty}. Practically, this potential is used for example to model stabilized colloidal systems and certain simple metals. From a theoretical perspective, its scale-free nature is of fundamental interest as it can be shown that all the system properties are a function of $\beta\rho^{n/3}$ only (in reduced units) \cite{hoover1971thermodynamic}. Additionally, liquids with interaction potentials that can be effectively approximated by it, also share this property to good approximation \cite{schroder2009hidden}. Here, we show data only for the state point given by ($k_BT/\epsilon = 1, \rho\sigma^3 = 3$, $n=5$). Because of its soft nature and high density, this system provides a good test for the theory and is complementary to the hard-sphere case.

Figure~\ref{fig:IPL} shows the most important results for this system. Again, the top panels (a, b) compare simulation data to thermodynamically inconsistent closures, and the bottom panels (c, d) compare to thermodynamically consistent ones. In the region $r<\sigma$, we show the bridge functions, and for $r>\sigma$ we show the error in the predicted $g(r)$. While we investigate a different state point and value of $n$, the features of the bridge function are similar to those obtained by \citet{llano1994bridge}. Out of the non-consistent closures, HNC performed the best, being known for its good performance for systems with soft potentials. The remaining non-consistent closures overestimate $-b(r<\sigma)$ significantly. Among the consistent ones, the more modern ERY, CJ, and CT closures reproduce $g(r)$ the most accurately, CJ outperforming the others regarding $b(r<\sigma)$. 

\begin{figure}[t]
    \centering
    \includegraphics[width=\linewidth]{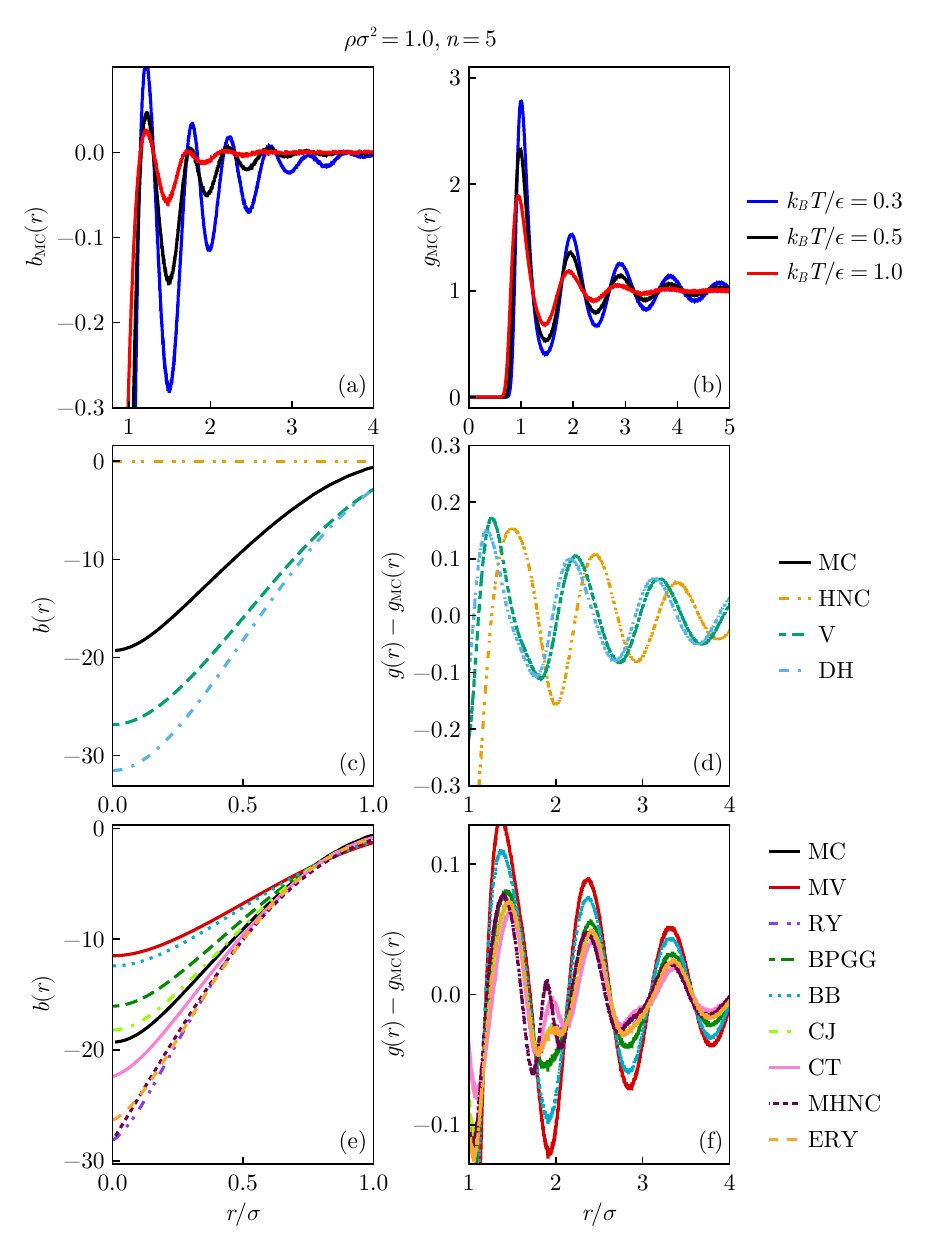}
    \caption{2D inverse power-law system: bridge function $b(r)$ and radial distribution function $g(r)$  compared to those obtained from Monte Carlo simulations. Panels (a) and (b) respectively show $b(r)$ and $g(r)$ obtained from computer simulations for three temperatures at $\rho\sigma^2=1.0$ and $n=5$. Panels (c--f) show the performance of various closures compared to the simulation data at $k_BT/\epsilon=0.5$.}
    \label{fig:IPL2d}
\end{figure}

As can be seen in panel (c), the simulation results for the bridge function suffered from bad statistics when $r\not\ll\sigma$ due to the high density. This does not affect the results for very small $r/\sigma$. Other state points that we investigated show similar behavior as shown in the figure. As $n$ increases, the results increasingly resemble those of the hard-sphere case.

\subsection{Inverse power law (2D)}

The phase diagram of the two-dimensional inverse power law system is similar to the hard disk system for large powers $n$, with both a discontinuous liquid-hexatic and a continuous hexatic-solid transition. However, as the value of $n$ is decreased beyond $n=6$, the transition between the liquid and the hexatic phase turns into a continuous transition \cite{kapfer2015two}. Near this transition, the correlation functions display behavior consistent with the Kosterlitz-Thouless-Halperin-Nelson-Young scenario of two-dimensional melting \cite{halperin1978theory, nelson1979dislocation, young1979melting}. 

On the liquid side of these transitions, the system behaves similarly to its three-dimensional counterpart. This is shown in Fig.~\ref{fig:IPL2d}, which summarizes our findings for this system at $n=5$. 
We find in agreement with the results for the three-dimensional inverse power law system, that among the thermodynamically inconsistent closures, the Verlet closure performs best regarding the zero-separation value $b(0)$, whereas HNC, V, and DH give results that are approximately equally accurate. Similarly, the thermodynamically consistent closures can be ranked the same in the two-dimensional case as they could in the three-dimensional system.

\subsection{Gaussian core model (3D)}\label{sec:GCM}

In addition to the inverse power-law and hard-sphere potentials, we consider the Gaussian core model as a third purely repulsive system \cite{stillinger1976phase}. This model displays qualitatively different behavior because the interaction potential is bounded, that is, it remains finite for any separation $r$. At low temperatures or high interaction strengths, it shows a re-entrant melting transition and a mean-field-like glass phase \cite{stillinger1976phase,  weber1981glass, prestipino2005phaseGCM, prestipino2005phase, ikeda2011glass}. Additionally, this model is of interest because in the high-density limit, the HNC closure becomes exact and the system reduces to an ideal gas \cite{louis2000mean}. In practice, it is used to model the effective interactions between polymer chains or similar objects \cite{louis2000can, lang2000fluid, likos2001effective, bolhuis2001accurate}.

\begin{figure}[t]
    \centering
    \includegraphics[width=\linewidth]{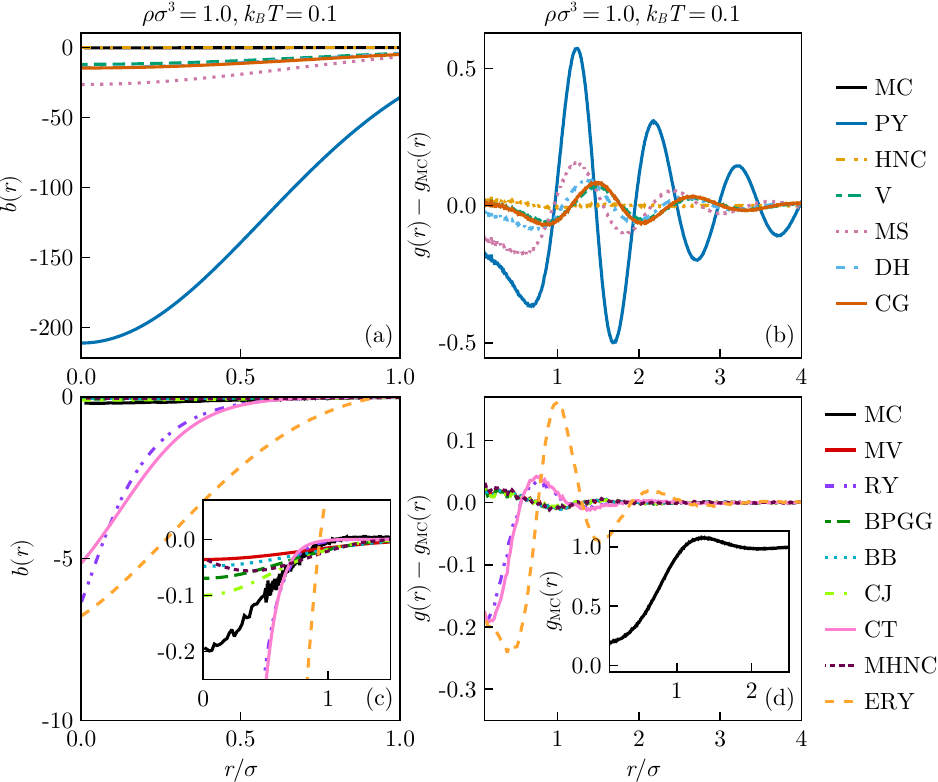}
    \caption{3D Gaussian-core model: accuracy of different closures compared to simulation data for $b(r)$ (a, c) and $g(r)$ (b, d). The inset of panel (c) shows a magnification of the data in the main panel, and that of (d) shows $g(r)$ obtained from simulations. In panel (a), the DH and CG curves overlap, making them hard to distinguish.}
    \label{fig:GCM}
\end{figure}

We compare the accuracy of the different closure relations for the Gaussian-core model in Fig.~\ref{fig:GCM}. We choose a state point ($\rho\sigma^3=1.0$, $k_BT/\epsilon = 0.1$) that is characterized by significant particle overlap and little positive correlations to contrast with the hard-sphere and inverse-power law systems. For this state point, the pair correlation function is shown in the inset of panel \ref{fig:GCM}(d). In particular, we find that $g(0) = 0.18$ and $g(r_\mathrm{peak}=1.37) = 1.08$. The results show a clear dichotomy between the different closures. Some show a serious overestimation of the magnitude of the bridge function, and thereby of the ordering and volume exclusion in the system. Out of the thermodynamically non-consistent closures only HNC performs well in predicting the bridge function. The radial distribution function is also reproduced best by HNC, but DH, V, and CG also provide reasonable predictions. Among the thermodynamically consistent closures, ERY, RY, and CT fail to predict particle overlap at zero-separation ($g(r\to0)>0$) and thus give qualitatively incorrect results. For the former two, this is not surprising, because both RY and ERY reduce to the PY equation at $r=0$.\cite{likos2001effective} All other thermodynamically consistent closures correctly do produce a very small $b(r\to0)$, slightly outperforming HNC (see the inset of panel \ref{fig:GCM}(c)). Among them, the CJ closure reproduces the Monte Carlo data best albeit not by much. Though not tested here, the random phase approximation closure $c(r) = -\beta u(r)$ is also known to be very successful for this model \cite{louis2000mean}. For tests of the closures at other state points for this system, see \citet{lang2000fluid, louis2000mean}, and \citet{likos2001effective}. It should be noted that the state point in question has a large influence on the accuracy rank order of the closures.

\begin{figure}[t]
    \centering
    \includegraphics[width=\linewidth]{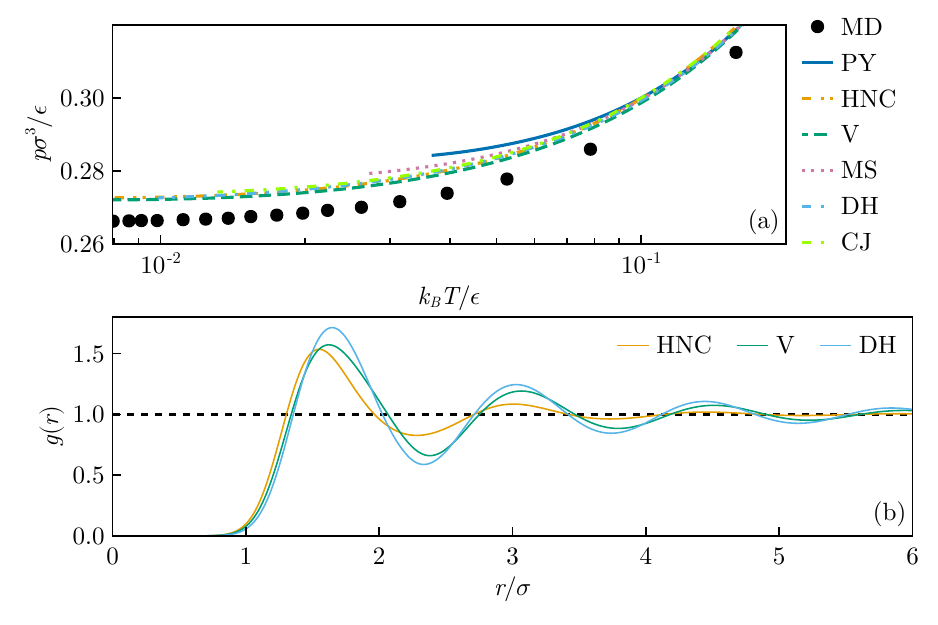}
    \caption{3D Gaussian-core model: virial pressure (a) obtained from various closures as a function of temperature at $\rho\sigma^3=0.33$ compared with Molecular Dynamics (MD) simulation data taken from \citet{ikeda2011thermodynamic} and radial distribution functions (b) at $k_BT/\epsilon=0.02$ and $\rho\sigma^3=0.33$ obtained from the HNC, V, and DH closure, which respectively correspond to pressures $p\sigma^3/\epsilon=0.2747$, $0.2740$, and $0.2749$.}
    \label{fig:GCM:p}
\end{figure}

In Fig.~\ref{fig:GCM:p}(a) we show the predicted pressure as a function of temperature for $\rho\sigma^3=0.33$. The symbols are simulation data taken from \citet{ikeda2011thermodynamic}, and the lines are our calculation for a small selection of the closure relations. We show that, as the temperature is lowered, each of the closure theories deviates from the simulation results. For sufficiently low temperatures, the predictions of some of the closures are nonphysical or fail to converge and are thus excluded from the figure. As noted before, it is important to realize that an accurately predicted pressure does not guarantee that the radial distribution function itself is predicted well. To illustrate this, we show the radial distribution functions for ($\rho\sigma^3=0.33,$ $k_BT/\epsilon=0.02$) as obtained with closures HNC, V, and DH. According to our numerics, they respectively predict pressures $p\sigma^3/\epsilon=0.2747$, $0.2740$, and $0.2749$, which deviate less than 0.5\%. However, the corresponding correlation functions, shown in Fig.~\ref{fig:GCM:p}(b), show significant differences, not only in the peak height but also in correlation length. The opposite does hold of course: if predictions for correlation functions overall improve in accuracy, the derived thermodynamic properties also become more precise. This gives justification for our focus on the accuracy of predictions of $g(r)$ and $b(r)$ directly, as opposed to predictions of thermodynamic properties (even though the latter is common practice). 

\subsection{Gaussian-core model (2D)}

\begin{figure}[t]
    \centering
    \includegraphics[width=\linewidth]{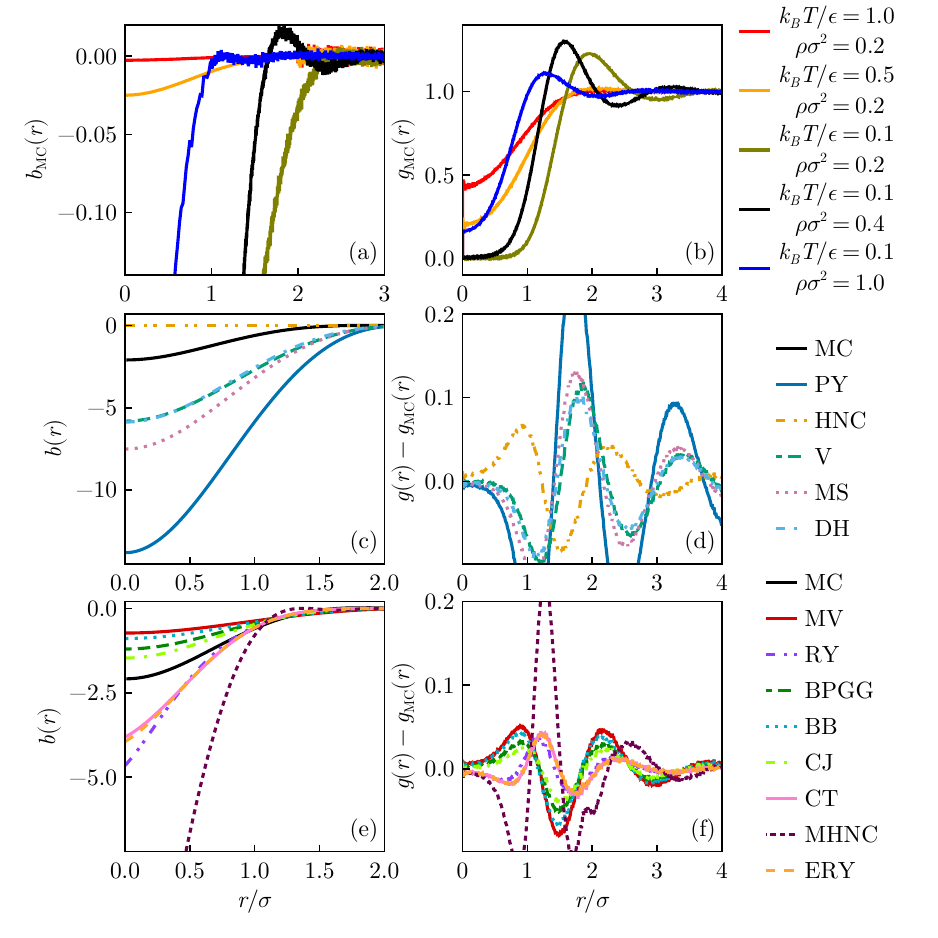}
    \caption{2D Gaussian-core model: bridge function and radial distribution function compared to those obtained from Monte Carlo simulations. Panels (a) and (b) respectively show $b(r)$ and $g(r)$ obtained from computer simulations for different state points. Panels (c--f) show the performance of various closures compared to the simulation data at $k_BT/\epsilon=0.1$, $\rho\sigma^2=0.4$, which corresponds to the black lines in panels (a,b)}
    \label{fig:gcm2d}
\end{figure}

The phase diagram of the two-dimensional Gaussian core model has been studied by \citet{prestipino2011hexatic} and by \citet{mendoza2024melting}. It features a continuous re-entrant two-step melting transition with associated critical temperature around $k_BT_c/\epsilon\approx0.01$.  
Figure \ref{fig:gcm2d} summarizes our results for this system. We focus our attention on the state point ($k_BT/\epsilon = 0.1, \rho\sigma^2 = 0.4$) which is far separated from the melting point. At this state point, the correlation functions show more structural features than our highlighted state point of the three-dimensional case, \textit{i.e.}, the first maximum of $g(r)$ is slightly higher, and the zero-separation values $g(0)$ and $b(0)$ are lower.

Due to the increased structural features, we find that the HNC closure, which gave excellent predictions in the 3D case, does not perform as well here. While it still outperforms the other non-consistent closures, the CJ and BPGG closures perform better here. Interestingly, the RY, CT, and ERY closures perform well here, as opposed to the three-dimensional case. This is a confirmation of the fact that their inaccuracy can be attributed to their inability to predict a positive $g(0)$ where appropriate for this model. 
 
\subsection{Lennard Jones (3D)}\label{sec:LJ}

\begin{figure}[t]
    \centering
    \includegraphics[width=\linewidth]{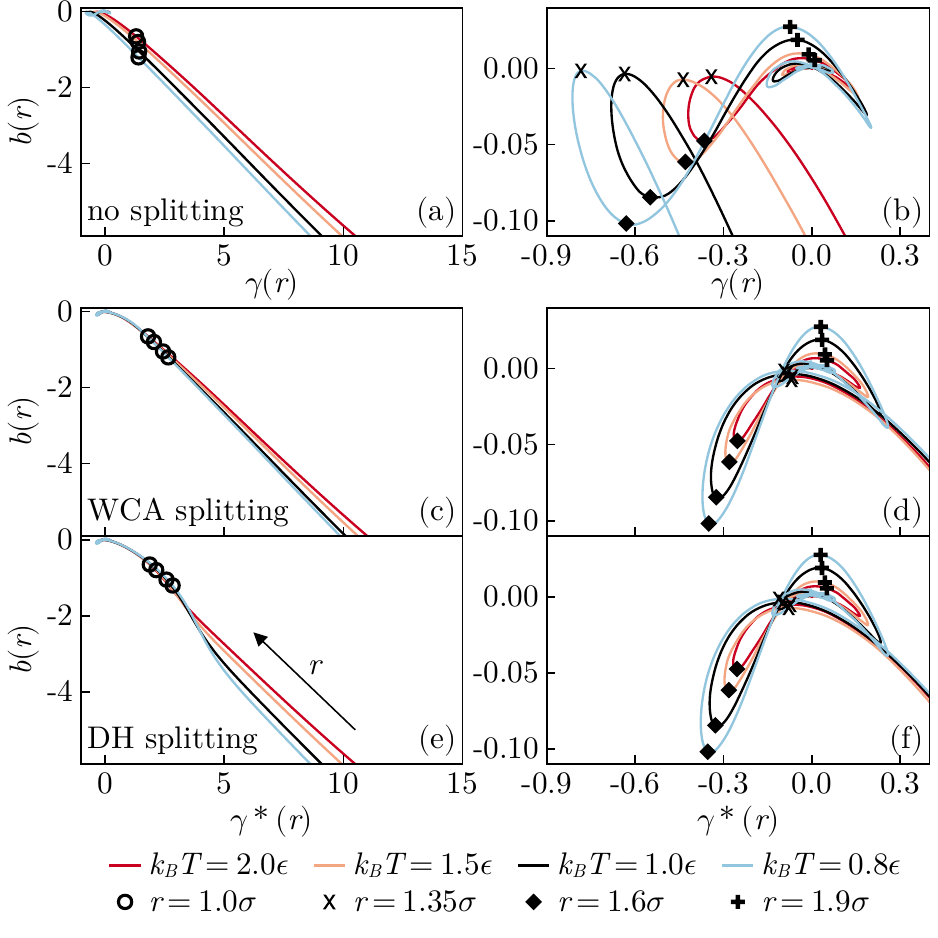}
    \caption{3D Lennard-Jones model: bridge function $b(r)$ as a function of the indirect correlation function $\gamma(r)$ (a, b) and renormalized indirect correlation function $\gamma^*(r) \equiv\gamma(r)-\beta u_\mathrm{LR}(r)$ (c--f) at different temperatures and number density $\rho\sigma^3=0.8$. The right-hand panels show magnifications of the left-hand ones. For clarity, the curves in the right panels have been smoothed by a local regression filter. Panels (c, d) show the $\gamma$-renormalization of \citet{weeks1971role} and (e, f) show that of \citet{duh1995integral}. The symbols show the locations where the parameter $r$ is equal to the value indicated in the legend.}
    \label{fig:LJ_bridge} 
\end{figure}

Up to this point, we have only considered purely repulsive systems. All liquids, however, comprise particles that attract each other. Attractive forces are therefore of vital importance in the theory of simple liquids. In particular, in relation to closure theory, they must be given special attention. The reason is visualized in Fig.~\ref{fig:LJ_bridge}(a, b), which shows the bridge function $b(r)$ for a Lennard-Jones system at different temperatures as a function of the indirect correlation function $\gamma(r)$. It is clear that any approach that approximates the bridge function locally as $b(r) \propto \gamma^2(r)$, for small $\gamma(r)$, fails qualitatively in the regime around $1<r/\sigma<1.6$. This issue was first addressed by \citet{madden1980mean}, though from a different motivation. They suggested that the indirect correlation function should be renormalized as $\gamma^*(r) = \gamma(r) - \beta u_\mathrm{LR}(r)$ to improve the accuracy. Indeed, with such a renormalization, the issue is partly resolved, as shown in Fig.~\ref{fig:LJ_bridge}(c, d) for the renormalization of \citet{weeks1971role} (WCA) and (e, f) for that of \citet{duh1995integral} (DH). This leads to a much more accurate prediction of the theories close to contact because $b(r)$ is locally much closer to being parabolic in $\gamma^*$ than in $\gamma$. Additionally, this renormalization causes the curves for the different state points to collapse in the region $1 < r/\sigma < 1.35$. This is seen as a second success of the renormalization procedure as it ensures that an accurate closure relation is state-point independent \cite{lee1995accurate}.

\begin{figure}[t]
    \centering
    \includegraphics[width=\linewidth]{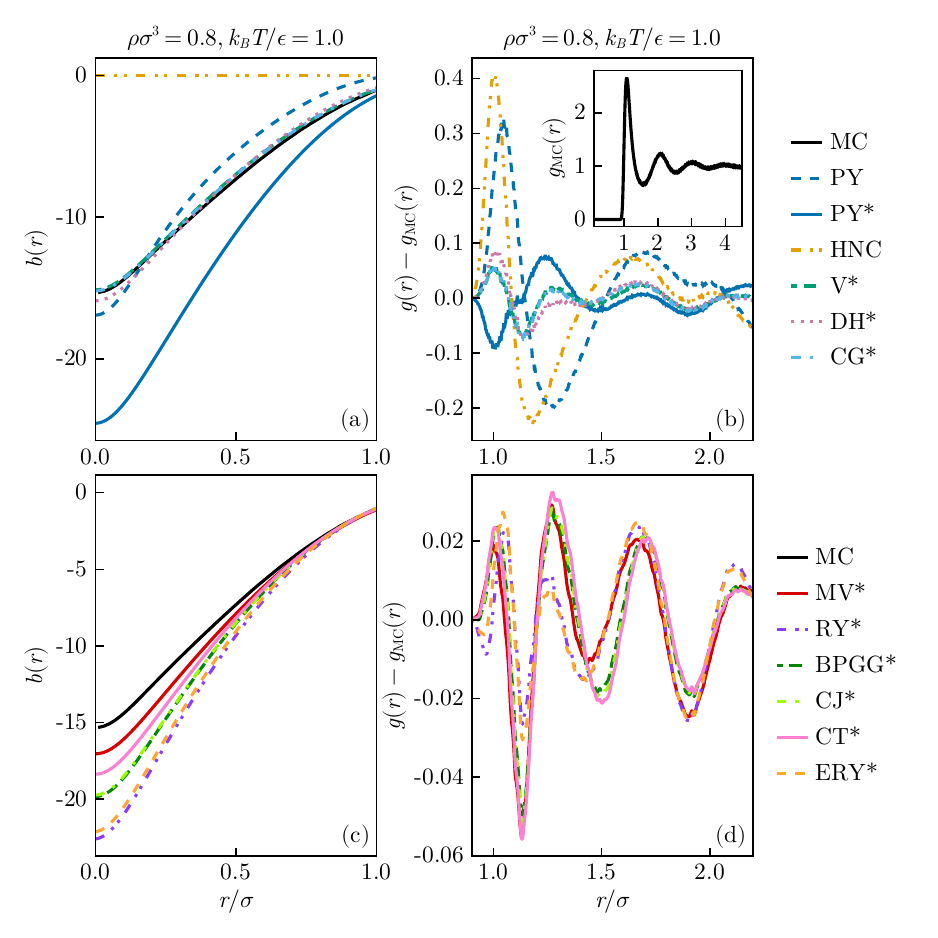}
    \caption{3D Lennard-Jones model: bridge function $b(r)$ (a, c) and radial distribution function $g(r)$ (b, d) ($\rho\sigma^3=0.8$, $k_B T/\epsilon=1.0$) compared to those obtained from Monte Carlo simulations. For context, the peak of $g(r)$ at this state point is $g(r_\mathrm{peak})\approx2.6$. The top panels show thermodynamically inconsistent closures, and the bottom panels show the consistent ones. The asterisks (*) in the closure acronyms indicate that the renormalized indirect correlation function $\gamma^*(r)$ is used in the closure as opposed to the standard indirect correlation function $\gamma(r)$. The inset of panel (b) shows the radial distribution function at this state point. }
    \label{fig:closures_LJ}
\end{figure}

Comparing the WCA and DH renormalization results in Figs.~\ref{fig:LJ_bridge}(c, e), we find for our state points that indeed the DH procedure yields a slightly better collapse of the curves than that of WCA near contact ($r\approx\sigma$). However, for distances $r/\sigma<1$, the curves fan out for DH, whereas the WCA results do not, making the latter easier to approximate or parameterize. Additionally, because the WCA splitting can be performed for any potential with a minimum, it is more generally applicable than the splitting procedure of \citet{duh1995integral}, which is specific to the Lennard-Jones liquid.  In the remainder of this section, therefore, we only consider the WCA renormalization.

The performance of various closure theories is compared in Fig.~\ref{fig:closures_LJ}. In the figure, closures indicated with an asterisk (*) use the renormalized indirect correlation function $\gamma^*(r) = \gamma(r) - \beta u_{\mathrm{LR}}(r)$. The long-ranged part of the potential is defined by the WCA splitting procedure. As can be seen in panels (a, b), the renormalization of the indirect correlation function in the closure relation for PY significantly improves the closure accuracy concerning the radial distribution function, while not necessarily improving the zero-separation value of the bridge function $b(0)$. Out of the thermodynamically inconsistent closures, the best performers are CG$^*$,  V$^*$, and DH$^*$.

The thermodynamically consistent closures all predict the radial distribution function roughly equally accurately (Fig.~\ref{fig:closures_LJ}d). The remaining error is caused by the inability to express $b(r)$ as a local function of $\gamma^*(r)$. For small distances $r<\sigma$, each of the closures overestimates the magnitude of the bridge function. For other comparisons of the accuracy of closure theories for the Lennard-Jones system, we refer the reader to e.g.~Refs. \citen{labik1983test, zerah1986self, lee1997potential, bomont2003new, bomont2008recent, ebato2016pressure, goodall2020machine, tsednee2019closure}.

\subsection{Lennard-Jones system (2D)}

\begin{figure}[t]
    \centering
    \includegraphics[width=\linewidth]{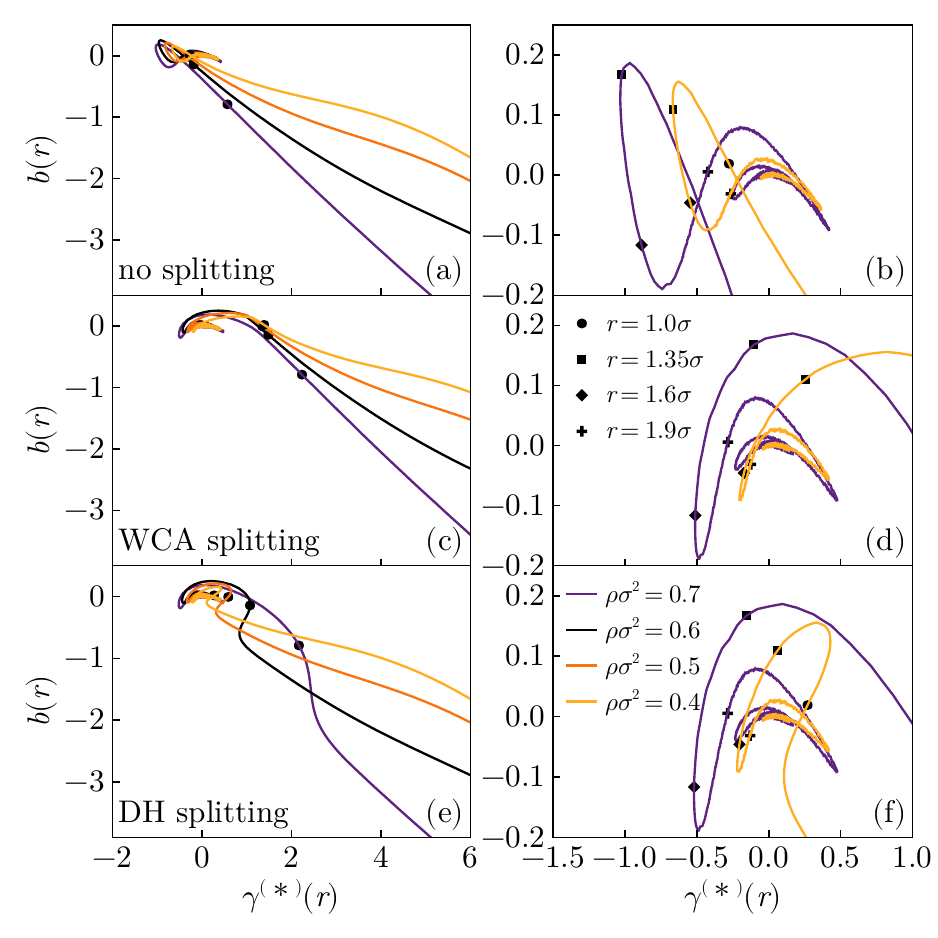}
    \caption{2D Lennard-Jones system: Duh-Haymet plots for different potential splitting schemes. The panels show the bridge function $b(r)$ against the (renormalized) indirect correlation function $\gamma^{(*)}(r)$ for different renormalization schemes: no splitting (a,b), WCA splitting (c,d) and DH splitting (e,f). Each panel on the left contains four different state points at $k_BT/\epsilon=0.6$. The right panels are magnifications of the left ones, showing only two out of four state points for clarity.}
    \label{fig:lj2dDH}
\end{figure}

\begin{figure*}[t]
    \centering
    \includegraphics[width=\linewidth]{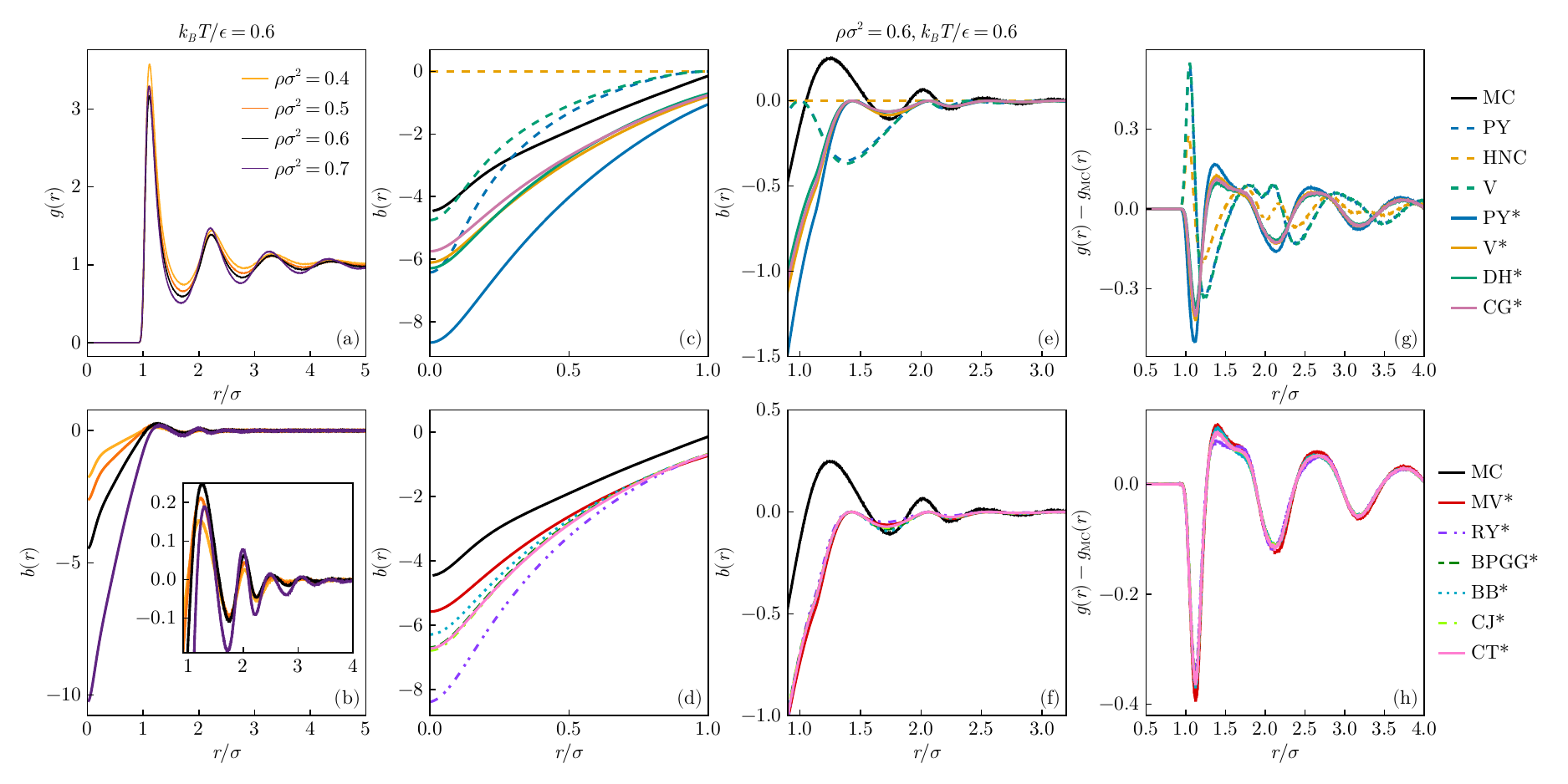}
    \caption{2D Lennard-Jones system: pair correlation and bridge function. Panels (a) and (b) respectively show the pair correlation function and bridge function as a function of $r$ for different densities. The other panels show the accuracy of various closure theories for the bridge function (c--f) and the pair correlation function (g, h). The closures indicated with an asterisk use the WCA renormalization scheme for the indirect correlation function. In panels (f) and (h) all closure predictions virtually overlap.}
    \label{fig:lj2dgr}
\end{figure*}

The liquid-gas critical point of the two-dimensional Lennard-Jones system is located at $\rho\sigma^2 = 0.488, k_BT/\epsilon=0.510$. \cite{phillips1981two, tsiok2022role} To avoid phase separation, we consider the supercritical regime $k_BT/\epsilon = 0.6$ at several different densities below the melting point (which is located at $\rho\sigma^2=0.82$ for this temperature \cite{phillips1981two, li2020phase}). One of the interesting questions to answer here is whether the renormalization schemes of the indirect correlation function, which were largely developed for the three-dimensional Lennard-Jones system, are also viable in its two-dimensional counterpart.

To answer this question, we draw the Duh-Haymet plots for this system in Fig.~\ref{fig:lj2dDH} for the WCA and DH splitting scheme of the pair potential. In the three-dimensional system these renormalization procedures have been shown to both make the bridge function $b(r)$ more amenable to local approximation in terms of $\gamma^*(r)$ as well as cause a collapse of the function for different state points. In its two-dimensional version, however, neither goals are achieved as shown in Fig.~\ref{fig:lj2dDH}. For short distances $r<\sigma$, we see that neither procedure reduces the fanning for large $\gamma$. At longer distances $r>1$, the failure of the renormalization procedure can be largely attributed to the non-negative peaks of $b(r)$, which are much more pronounced here than for the other systems we studied in this work. These can also be seen clearly in Fig.~\ref{fig:lj2dgr}(b) which shows the $b(r)$ as a function of $r$. 

Fig.~\ref{fig:lj2dgr}(c--h) shows the predictive performance of different closure theories for this system at $\rho\sigma^2=0.6$. We show the accuracy of the predicted bridge function (c--f) as well as that of the predicted radial distribution function $g(r)$. The predictions show that the closures can be classified into three clusters based on performance: renormalized closures, nonrenormalized closures, and a cluster containing only the HNC closure. Within each cluster, the results are almost indistinguishable in the $r>\sigma$ regime. In the $r<\sigma$ regime, the normalized closures perform slightly better qualitatively.

\section{Conclusion}

We have presented an accuracy comparison between integral equation theories based on the Ornstein-Zernike theory across different model systems. We have consistently separated thermodynamically consistent theories from non-consistent ones. The former have the advantage of typically providing more accurate correlation functions and thermodynamic quantities, while the latter are computationally simpler to implement and solve. Our results show that there is no closure theory that performs best across all systems. Nevertheless, some closure relations performed better, overall, than others. In particular, the non-consistent closure presented in \citet{verlet1980integral} (V) stood out because it robustly performed well for all our benchmark systems, indicating that it could suitably serve as a first guess for use in unfamiliar systems. Among the consistent closures, the differences in accuracy are smaller. The Modified Verlet (MV) closure and the closure by \citet{charpentier2001exact} (CJ) both performed excellently in most cases. For steeply repulsive potentials, MV proved more accurate, while CJ showed superiority for softer, longer-ranged potentials. 

Across all systems studied, the well-known Percus-Yevick equation performed worse than any among the V, DH, and MS closures. This comparative underperformance of the well-known and highly used Percus-Yevick equation, especially in comparison to other non-consistent closures, highlights the importance of considering alternatives when accuracy takes precedence over the necessity for analytical solutions.

By comparing our analysis in two dimensions with that in three, we conclude that the relative accuracy of the closure theories seems relatively stable across dimensions. This is fortunate since there is relatively little data available for the performance of different closures in two-dimensional systems. One exception is that the renormalization procedure of the indirect correlation function that is often employed in systems with attractive potential tails seems much less effective in two dimensions than in three.

Our study has targeted only single-component fluids with uncharged, rotationally symmetric two-body interaction potentials. It is unlikely that all our conclusions hold when these restrictions are relaxed, but we leave their investigation to further research. 

To simplify both the selection of a suitable closure relation for new model systems and the evaluation and dissemination of novel closure relations, we have packaged and documented our code to solve the Ornstein-Zernike equation in an open-source repository \cite{OZjl}, which is straightforward to install and use from interactive programming languages such as Python and Julia. It is capable of solving the equations for single-component systems and mixtures at any spatial dimension for a given (user-defined) closure and radial pair potential.

\section*{Acknowledgements}
The Dutch Research Council (NWO) is acknowledged for financial support through a Vidi grant.

\section*{Data availability}
The data that support the findings of this study are available from the corresponding author upon reasonable request. The software is published at Ref.~\citen{OZjl}.

\section*{Author contributions}
\textbf{IP}: Conceptualization, Data Curation, Software, Writing.\\
\textbf{LMCJ}: Conceptualization, Writing (Review and Editing), Supervision, Funding Acquisition.

\bibliography{apssamp}

\appendix

\section{Accurate Fourier-Bessel transforms}\label{app:quad}
Here we present accurate quadrature for the radial Fourier transform. These are only used for inversion of the Ornstein-Zernike equation, given measurements for $g(r)$. They are not suitable for use in iterative solution methods for the integral equation because their orthogonality is not enforced.

For three dimensions, let 
\begin{equation}
    I(k) = \int_0^\infty \mathrm{d}x  \sin(kx) x f(x).
\end{equation}
To approximate $I(k)$, we discretize $x$ to the grid $x_i = (i - 1/2)h$, where $i = 1,\ldots,M$, and $h$ can be chosen such that $f(Mh)\approx 0$. Now, assuming $f(x)$ is constant in the interval $x_i - h/2 < x < x_i + h/2$, and with a value of $f_i = f(x_i)$, we have \cite{kolafa2002bridge}
\begin{align}
    I(k) &= \sum_{i=1}^M  \int_{x_i-h/2}^{x_i+h/2} \mathrm{d}x  \sin(kx) x f(x)\\
    &\approx \sum_{i=1}^M  f_i \int_{x_i-h/2}^{x_i+h/2} \mathrm{d}x  \sin(kx) x \\
    &= \frac{1}{k^2} \sum_{i=1}^M  f_i \Big[2\sin\left(\frac{hk}{2}\right)\left(\cos(k x_i) + x_i k \sin(k x_i)\right)\nonumber\\&\qquad-hk\cos\left(\frac{hk}{2}\right)\cos(kx_i) \Big],
\end{align}
which is essentially the midpoint rule with a $x\sin(kx)$ weight kernel.

Alternatively, for the two-dimensional case, let

\begin{equation}
    I(k) = \int_0^\infty \mathrm{d}x  J_0(kx) x f(x).
\end{equation}
With the same discretization, we find
\begin{align}
    I(k) &= \sum_{i=1}^M  \int_{x_i-h/2}^{x_i+h/2} \mathrm{d}x  J_0(kx) x f(x)\\
    &\approx \sum_{i=1}^M  f_i \int_{x_i-h/2}^{x_i+h/2} \mathrm{d}x  J_0(kx) x \\
    &= \frac{1}{k} \sum_{i=1}^M  f_i \bigg[\left(x_i+\frac{h}{2}\right)J_1\left(kx_i+\frac{hk}{2}\right)\nonumber\\&\qquad -\left(x_i-\frac{h}{2}\right)J_1\left(kx_i-\frac{hk}{2}\right) \bigg].
\end{align}

\end{document}